\documentclass[preprint,aps,showpacs,preprintnumbers,amsmath,amssymb,superscriptaddress,floatfix,longbibliography]{revtex4-1}

\usepackage{graphicx}
\usepackage[ansinew]{inputenc}
\usepackage{array}
\usepackage{color}
\usepackage{amsmath}
\usepackage{amsxtra}
\usepackage{amstext}
\usepackage{amssymb}
\usepackage{latexsym}

\usepackage{sidecap}
\usepackage{floatflt}

\begin{document}

\title{Theory of optically controlled anisotropic polariton transport in semiconductor double microcavities}

\author{S.M.H. Luk}
\affiliation{Department of Physics, University of Arizona, Tucson, AZ 85721, USA}

\author{P. Lewandowski}
\affiliation{Physics Department and Center for Optoelectronics and Photonics Paderborn (CeOPP), Universit\"at Paderborn, Warburger Strasse 100, 33098 Paderborn, Germany}

\author{N.H. Kwong}
\affiliation{College of Optical Sciences, University of Arizona, Tucson, AZ 85721, USA}

\author{E. Baudin}
\affiliation{Laboratoire Pierre Aigrain, D\'epartement de physique de
l'ENS, Ecole normale
sup\'erieure, PSL Research University, Universit\'e Paris Diderot,
Sorbonne Paris Cit\'e, Sorbonne
Universit\'es, UPMC Univ. Paris 06, CNRS, 75005 Paris, France}

\author{O.\ Lafont}
\affiliation{Laboratoire Pierre Aigrain, D\'epartement de physique de
l'ENS, Ecole normale
sup\'erieure, PSL Research University, Universit\'e Paris Diderot,
Sorbonne Paris Cit\'e, Sorbonne
Universit\'es, UPMC Univ. Paris 06, CNRS, 75005 Paris, France}

\author{J. Tignon}
\affiliation{Laboratoire Pierre Aigrain, D\'epartement de physique de
l'ENS, Ecole normale
sup\'erieure, PSL Research University, Universit\'e Paris Diderot,
Sorbonne Paris Cit\'e, Sorbonne
Universit\'es, UPMC Univ. Paris 06, CNRS, 75005 Paris, France}

\author{P.T. Leung}
\affiliation{Department of Physics and Institute of Theoretical Physics, The Chinese University of Hong Kong, Hong Kong SAR, China}

\author{K.P. Chan}
\affiliation{Department of Physics and Institute of Theoretical
Physics, The Chinese University of Hong Kong, Hong Kong SAR, China}

\author{M. Babilon}
\affiliation{Physics Department and Center for Optoelectronics and Photonics Paderborn (CeOPP), Universit\"at Paderborn, Warburger Strasse 100, 33098 Paderborn, Germany}



\author{S. Schumacher}
\affiliation{Physics Department and Center for Optoelectronics and Photonics Paderborn (CeOPP), Universit\"at Paderborn, Warburger Strasse 100, 33098 Paderborn, Germany}
\affiliation{College of Optical Sciences, University of Arizona, Tucson, AZ 85721, USA}

\author{R. Binder}
\affiliation{College of Optical Sciences, University of Arizona, Tucson, AZ 85721, USA}
\affiliation{Department of Physics, University of Arizona, Tucson, AZ 85721, USA}

\begin{abstract}
Exciton polaritons in semiconductor microcavities exhibit many fundamental physical effects, with some of them amenable to being controlled by external fields.
The polariton transport is affected by the polaritonic spin-orbit interaction, which is caused by the splitting of transverse-electric and transverse-magnetic (TE-TM) modes.
This is the basis for a polaritonic Hall effect, called optical spin Hall effect (OSHE), which is related to the formation of spin/polarization textures
in momentum space, determining anisotropic ballistic transport, as well as related textures
in real space.
 Owing to Coulombic interactions between the excitonic components of the polaritons,
optical excitation of polaritons can affect the OSHE.
 We present a theoretical analysis of the OSHE and its optical control in semiconductor double microcavities, i.e.\ two optically coupled cavities,
which are particularly well suited for the creation of polaritonic reservoirs that affect
the spin-texture-forming polaritons. The theory is formulated in terms of a set of
double-cavity spinor-polariton Gross-Pitaevskii equations.
 Numerical solutions feature, among other things, a controlled rotation of
the spin texture in momentum space. The theory also allows for an identification of the effective magnetic field component that determines the optical control in
phenomenological pseudo-spin models in
terms of exciton interactions and the polariton density in the second lower polariton branch.
 \end{abstract}



\today

\maketitle


\section{Introduction}
\label{sec:intro}

Exciton polartions in semiconductor microcavities have been extensively studied for many years
\cite{savona-etal.94,gonokami-etal.97,fan-etal.98,baars-etal.00,kwong-etal.01prl,kwong-etal.01prb},
 in part because
many of the intriguing features of cavity polaritons are associated with their unique dispersion and small effective mass.
Parametric amplification of polaritons, which utilizes the so-called ``magic angle'' (the inflection point on the LPB) was observed.
Here, a weak probe beam in normal incidence experiences large amplification when a strong beam pumps the polaritons at this angle
(for example \cite{Savvidis2000,baars-etal.00,Huang2000,Ciuti2000,Stevenson2000,houdre-etal.00,Saba2001,Whittaker2001,%
Ciuti2003,Savasta2003,Savasta2003b,langbein.04,Baumberg2005,Keeling2007}; for reviews see for example
Refs.~\cite{Ciuti2003,Baumberg2005,Keeling2007,sanvitto-timofeev.12}).
The small polariton mass has also allowed for the successful observation of polariton Bose condensates
\cite{deng-etal.02,Kasprzak2006,Balili2007,utsunomiya-etal.08,deng-etal.10,snoke-littlewood.10,moskalenko-snoke.00}.

An important aspect of much of the aforementioned effects is the interaction between polaritons
(e.g.\ \cite{%
jahnke-etal.97,%
suzuura-etal.98,%
shirane-etal.98,%
quochi-etal.98,%
fan-etal.98,%
khitrova-etal.99,%
okumura-ogawa.00,%
svirko-gonokami.00,%
neukirch-etal.00prl,%
houdre-etal.00,%
combescot-etal.07,%
pilozzi-etal.10,%
combescot-shiau.16}).
A detailed
non-perturbative (in the Coulomb interaction) T-matrix analysis of
two-exciton correlations in GaAs quantum wells, that
fully determines the
nonlinear optical response in the coherent third-order (or $\chi ^{(3)}$)
regime, was presented in Refs.\
\cite{kwong-binder.00,takayama-etal.02,takayama-etal.04}.
 This work includes the full vectorial dependence of the third-order
susceptibility tensor and a biexcitonic resonance in the two-exciton T-matrix. In the T-matrix calculations
both non-zero wavevector wave functions as well at intermediate exciton states with non-zero angular momentum are included. This work also provides a unifying foundation of nonlinear polariton physics in that it clarifies the interaction's
spin dependence, energy dependence, and 2D characteristics (e.g.\ the logarithmic approach to zero at the continuum threshold).
The biexcitonic resonance discussed in \cite{takayama-etal.02,schumacher-etal.07prb}
has recently been related to  a Feshbach resonance \cite{takemura-etal.14} in microcavity
polaritons. Further work on
the third-order nonlinear response and exciton-exciton interactions in GaAs quantum wells
includes Refs.\
\cite{%
axt-stahl.94,%
maialle-sham.94,%
lindberg-etal.94b,%
oestreich-etal.98,%
axt-mukamel.98,%
sieh-etal.99,%
koch-etal.01,%
mayer-etal.94b,%
schaefer-etal.96,%
bartels-etal.97,%
kner-etal.99,%
meier-etal.00,%
neukirch-etal.00prb,%
bolton-etal.00,%
brick-etal.01,%
axt-etal.01b,%
langbein-etal.01,%
lecomte2014,
okumura-ogawa.01}.

In addition to exciton-exciton interactions, which lead to a polariton-polariton interaction,  semiconductor microcavities
have an inherent polariton-spin orbit interaction, which can be compared to spin-orbit interaction in other optical systems, for example
the plasmonic spin Hall effect \cite{xiao-etal.15}, and
spin orbit interactions of light
\cite{%
liberman-zeldovich.92,%
brasselet-etal.09,%
rodriguez-herrera-etal.10,%
yin-etal.13,%
bliokh-etal.15%
}.
The physical origin of the polariton-spin orbit interaction is the
transverse-electric and transverse-magnetic (TE-TM)
 splitting, which
can be described in terms of an effective magnetic field, and which in turn
  gives rise to a polaritonic spin Hall effect,
 which is called optical spin Hall effect (OSHE)
 \cite{kavokin-etal.05,%
leyder-etal.07,%
langbein-etal.07,%
manni-etal.11,%
maragkou-etal.11,%
kammann-etal.12}.
Since polaritons with different in-plane wave vectors
$\textbf{k}$
experience different effective magnetic fields
$\textbf{B}(\textbf{k})$,
an  isotropic distribution of polaritons on a ring in wave vector space
can lead to an anisotropic polarization texture or pattern, both in real  and momentum (or wave vector) space.
Such polarization/spin textures
have been found for excitations of linearly  and circularly  polarized polaritons in Ref.\
\cite{leyder-etal.07} and \cite{kammann-etal.12}, respectively
  (structurally similar polarization/spin textures are also present in different physical systems, e.g.\
\cite{hielscher-etal.97,schwartz-dogariu.08}).
The OSHE in wave vector space  corresponds to that seen experimentally in far-field observations, which are particularly important in possible photonics or spinoptronics \cite{shelykh-etal.04,shelykh-etal.10} applications.
The far field contains information on the (in this case anisotropic) ballistic polariton transport, as it is an image of the spin-dependent polariton density in momentum (or wave vector) space.

\begin{figure}
\centerline{\includegraphics[scale=0.6,angle=-00,trim=00 0 00 00,clip=true]{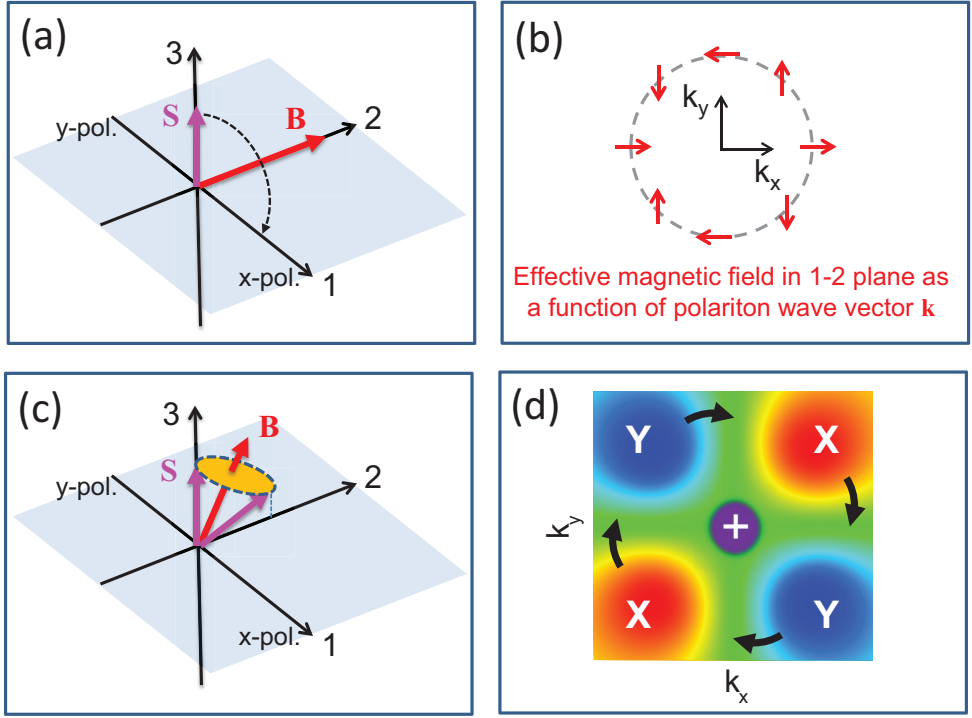}}
\caption{
(a) Sketch of pseudo-spin space and torque action when $\textbf{B}$ has only a positive $B_2$ component (corresponding to polariton wave vector $\textbf{k}$ along the $45^{\circ}$ direction.
(see (b)).
(b) Sketch of the effective magnetic field vectors (red) for different directions of the polariton wave vector $\textbf{k}$.
(c) Similar to (a), but for a $\textbf{B}$ with a non-zero $B_3$ component.
(d) Sketch of the steady state $S_1$ component of the pseudo-spin vector as a function of the polariton wave vector $\textbf{k}=(k_x,k_y)$. Red (blue) color corresponds to predominant x (y) polarization. The circularly (``+'') polarized pump at $\textbf{k}=0$ (normal incidence) is also indicated. The black arrows indicate the rotation of the pattern that occurs when $B_3$ is non-zero and positive.
}
\label{fig:steady-state-torque}
\end{figure}

A system that is well-suited for the control of the polariton-spin transport and the OSHE is a semiconductor double microcavity.
 The theory of polaritons in coupled multiple microcavities has been studied in Ref.~\cite{panzarini-etal.99prb},  a demonstration of optical parametric oscillation (OPO) in a triple microcavity has been reported in \cite{diederichs-etal.06}, and
generation of entangled photons in coupled microcavities has been proposed in \cite{portolan-etal.14}. Furthermore, the double cavity
has been used recently in an experimental observation of the controlled OSHE \cite{lafont-etal.17}.
In that work, the experimental observations are explained with the help of a
 simple pseudo-spin model \cite{kavokin-etal.05,leyder-etal.07}. The pseudo-spin vector $\textbf{S}$ is given in terms of
the Stokes parameters formed by the polariton wave functions  (its precise definition will be given below).

For our initial discussion it is sufficient to note that the pseudo-spin model highlights the effect of the spin-orbit interaction on quantities that are
 readily observable in the far field, namely the components of $\textbf{S}(\textbf{k})$, where $\textbf{k}$ is the polariton's in-plane wave vector.
The evolution of $\textbf{S}$ is given by an equation that includes a torque term, which has the standard  $ \textbf{B}(\textbf{k} ) \times \textbf{S}(\textbf{k}) $ form, found in many magnetic systems:
\begin{equation}
 \left. \frac{\partial}{\partial t}\textbf{S}(\textbf{k},t)  \right|_{\rm torque} = \frac{1}{\hbar}\left(\textbf{B}(\textbf{k}) \times \textbf{S}(\textbf{k},t)\right),\label{pseudospin.equ}
\end{equation}
with $\textbf{B}(\textbf{k}) = \left(\Delta_k\cos(2\phi_k),\Delta_k\sin(2\phi_k),B_3 \right)$. Here $\Delta_k$ is the TE-TM splitting at $k=|\textbf{k}|$. It will become clear from the definition of $\textbf{S}(\textbf{k})$  given below that
positive (negative) $S_1$ corresponds to dominant x (y) linear polarization components of the polariton fields, and positive (negative) $S_3$ to dominant ``+'' (``-'') circular polarizations.
Additional  details of the torque model in the context of the optical control of the OSHE, including the complete equation of motion for   $\textbf{S}(\textbf{k},t) $, are given in \cite{lafont-etal.17}.

If we assume ``+'' circularly polarized excitation and consider the example of a wave vector in the diagonal direction, $\textbf{k} \propto (1,1,0)$, in other words the polar angle $\phi_k $ of the wave vector is $45^{\circ}$, then it turns out that the effective magnetic field has only a positive $B_2$ component\cite{kavokin-etal.05,leyder-etal.07}. Initial circularly polarized polaritons will then experience a torque as schematically indicated in Fig.\ \ref{fig:steady-state-torque}a, which rotates $\textbf{S}$ towards positive $S_1$, i.e.\ towards x-polarization, until it settles to steady state (in this case in the 1-3 plane).
At $\phi_k = -45^{\circ}$, the negative $B_2$ component of $\textbf{B}$ would favor y-polarization, and the magnetic field directions as a function of wave vector directions are schematically shown in Fig.\ \ref{fig:steady-state-torque}b. In the low-excitation regime, in which polaritonic interactions can be neglected, the effective magnetic field is restricted to the 1-2 plane.
The possible control of the effective magnetic field is based on polaritonic interactions. For the case of conventional single-cavity structures,
it has been pointed out that a $B_3$ component can arise from polariton populations \cite{lagoudakis-etal.02,shelykh-etal.10}. With a non-zero $B_3$ component, the polariton-spin dynamics related to the torque-like force is more complicated, as schematically indicated in Fig.\ \ref{fig:steady-state-torque}c.
In Ref.\ \cite{lafont-etal.17} it was found that the double-cavity allows for a well-controlled rotation of the polaritonic spin/polarization texture in wave vector space, as indicated by the black arrows in Fig.\ \ref{fig:steady-state-torque}d.

In the following, we develop a microscopic framework for the analysis of the spin/polarization textures in real and wave vector space, we analyze the nonlinear case including polaritonic ineractions, and we identify the $B_3$ component of the effective magnetic field in terms of microscopic quantities such as polariton wave functions and excitonic T-matrices.

\section{Polaritons in single and double microcavities}

Before presenting more details of the microscopic theory in the following sections, let us briefly summarize key aspects of semiconductor polaritonic microcavities.
In their simplest form, such cavities consist of one (or several) semiconductor quantum wells between two mirrors (such as distributed Bragg reflectors, DBR). Excitons
interact with the light field inside the cavity to form exciton polaritons.


\begin{figure}
\includegraphics[scale=0.8]{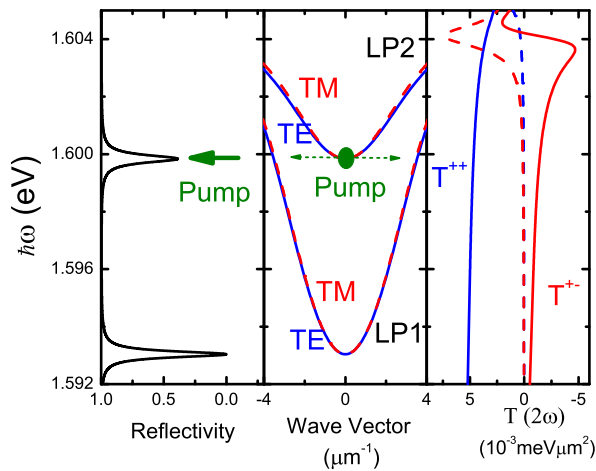}
\caption{
The frequency-dependent reflectivity, dispersion relations of the two lower polariton branches, and real (solid lines) and imaginary (dashed lines) parts of exciton-exciton interactions
 $T^{++}$ and $T^{+-}$ from Ref.\ \cite{takayama-etal.02}.
 The branch LP2 allows the normal incidence pump to enter the cavity.
 The LP2 polaritons act as a source for LP1 polaritons, as indicated by the dashed horizontal arrows, and simultaneously as a polariton reservoir that controls the orientation of the OSHE pattern.
 The  splitting between TE and TM polaritons, particularly large in our double-cavity structure,
  results in spin-orbit coupling via an effective magnetic field\cite{kavokin-etal.05}.
  Modification of the effective magnetic field is enabled through optical pumping and the resulting spin-dependent interaction between polaritons,
  which in turn is based on the underlying exciton-exciton interaction.
}
\label{fig:cavity-dispersion}
\end{figure}

The excitonic component of the polariton provides a strong Coulombic spin-dependent exciton-exciton interaction which is conceptually similar to the Heitler-London model for chemical bonds in hydrogen molecules,
where the sum (difference) of direct and exchange interactions determine the singlet (triplet) molecular energies.
For excitons, the results in interactions $T^{++}$ ($T^{+-}$) are different if the two excitons have same (opposite) circular polarization.
For quantum well excitons, these interactions have been
 quantitatively evaluated in Ref.\
 \cite{takayama-etal.02}, and  for polaritons they result in a spin and energy-dependent interaction\cite{schumacher-etal.07prb}.
As mentioned above, these interactions have been found to be important in  spin-dependent optical nonlinearities
 (e.g.\ Ref.\ \cite{kwong-etal.01prl,lecomte2014}),
Bose-Einstein condensates (for a review, see e.g.\ Ref.\ \cite{deng-etal.10}),
spin-dependent optoelectronic devices
(e.g.\ Ref.\ \cite{shelykh-etal.04,shelykh-etal.10}), and polaritonic Feshbach resonances\cite{takemura-etal.14}.
In the present context, they modify the effective spin-orbit interaction by modifying the effective magnetic
field\cite{shelykh-etal.10}, and hence the spin and polarization texture of the polariton field in real and wave vector space\cite{flayac-etal.13prl}.

When discussing Fig.\  \ref{fig:steady-state-torque}c, we noted that
 a circularly (``+'') polarized pump beam results in the tilting of the effective magnetic field $\textbf{B}(\textbf{k})$
 out of the 1-2 plane in pseudo-spin space, and hence can lead to a modification of the spatial polarization texture of the polaritons' linear polarization  components,
Fig.\  \ref{fig:steady-state-torque}d.
 However, a simple implementation of such an optical pump scheme,
 which involves only a single spatial component and a quasi-monochromatic pump acting both as the source of the polaritons involved in the OSHE and
 their external control,
 is difficult in conventional single-cavity designs. A double-cavity that contains two lower polariton branches can overcome those difficulties. A double-cavity consists of two sets of quantum wells embedded in two cavities \cite{lafont-etal.17, ardizzone-etal.13}. The two cavities are coupled with each other, and within each cavity, the cavity field is coupled with the excitons of the quantum wells. This forms four polariton branches, with two upper and two lower polariton branches.

The reflectivity spectrum of a double-cavity  is shown in
Fig.\
\ref{fig:cavity-dispersion}.
In contrast to an ordinary single-cavity design, in the double-cavity system the reflectivity spectrum exhibits two minima.
The upper of the two lower polariton branches, LP2, yields a reflectivity window that allows a normal-incidence monochromatic pump to enter the cavity, creating LP2 polaritons.   In the following, the upper polaritons are neglected as they are strongly detuned from the pump frequency.

%

The pump exciting LP2 polaritons also creates
a small population of LP1 polaritons, mostly indirectly via elastic Rayleigh scattering,
 but, depending on the pump beam profile, possibly also through overlap of the pump's spatial frequencies with the LP1 dispersion.
%
 %
 Due to the underlying exciton-exciton interactions between LP1 polaritons on the elastic circle and LP2 polaritons close to $k=0$,
 shown in
 Fig.\ \ref{fig:cavity-dispersion},
 the pump can induce a shift of the LP1 frequencies.
As we will show below, the shift is different for ``+'' and ``-'' polarized LP1 states, which is important for the control of the OSHE.
For typical parameters, such as those used in Ref.\ \cite{lafont-etal.17},
 the shifts are substantial (of order 0.5meV). We stress that
   in the present study we stay
  below the optical parametric oscillation (OPO) threshold. This is in contrast to our earlier work on polaritonic pattern formation \cite{ardizzone-etal.13}, which focused
  on a regime above OPO threshold.

\section{Double-cavity spinor Gross-Pitaevskii equations}

Spinor-valued polaritons in a double cavity are represented by wave functions $\psi^{\pm}_m(\textbf{r},t)$ (where $m$ is LP1 or LP2),
which obey  driven polaritonic Gross-Pitaevskii equations.
The Gross-Pitaevskii equations for the double cavity  are derived in this section.
 For the single-cavity case they can be found, for example, in Ref.\ \cite{liew-etal.08}.
 In Ref.\ \cite{ardizzone-etal.13} the double-cavity is described in terms of the
 more fundamental equations for the excitonic interband polarization $p$ and the electric field amplitude $E$.
 However, such a general description makes it difficult to elucidate the nature of the anisotropic ballistic polariton transport and
 in particular interaction effects that give rise to an effective magnetic field. The double-cavity polaritonic Gross-Pitaevskii equations
  that we derive in the following allow us to identify the effective magnetic field with the polaritonic wave functions and interactions.
  Since there is a well-defined relation between the polariton picture and the underlying $p$ and $E$ amplitudes, it is then possible
  to provide a quantitative relation between the effective magnetic field and the power of the incident light field.

The equations of motion for excitons and cavity photon fields in semiconductor quantum well double microcavities  were given in Ref.\ \cite{ardizzone-etal.13} in a $\textbf{k}$-space formulation. They consist of equations for the electric field amplitude at the position of the quantum well within the quasi-mode approximation and equations for the exciton (specifically the 1s heavy-hole exciton) component of the interband polarization. In a real-space formulation they read
\begin{eqnarray}
i \hbar \dot{E}^{\pm}_{1} &=& H_C E^{\pm}_{1} + H_{\pm} E^{\mp}_{1} - \Omega_C E^{\pm}_{2} -\Omega_X p^{\pm}_{1} + R_{\text{pump},1}^{\pm} \label{E1.equ}   \label{equ:E-dot-pm-1}  \\
i \hbar \dot{E}^{\pm}_{2} &=& H_C E^{\pm}_{2} + H_{\pm} E^{\mp}_{2} - \Omega_C E^{\pm}_{1} -\Omega_X p^{\pm}_{2} + R_{\text{pump},2}^{\pm} \label{E2.equ} \\
i \hbar \dot{p}^{\pm}_{1} &=& H_X p^{\pm}_{1} - \Omega_X (1-2A_{\text{PSF}} |p^{\pm}_{1}|^2)E^{\pm}_{1} \nonumber \\ &&+ T^{++} |p^{\pm}_{1}|^2 p^{\pm}_{1} + T^{+-} |p^{\mp}_{1}|^2 p^{\pm}_{1} \label{p1.equ} \\
i \hbar \dot{p}^{\pm}_{2} &=& H_X p^{\pm}_{2} - \Omega_X (1-2A_{\text{PSF}} |p^{\pm}_{2}|^2)E^{\pm}_{2} \nonumber \\ &&+ T^{++} |p^{\pm}_{2}|^2 p^{\pm}_{2}   + T^{+-} |p^{\mp}_{2}|^2 p^{\pm}_{2} \label{p2.equ}
\end{eqnarray}
with $H_C = -\frac{\hbar^2}{4} \left(\frac{1}{m_{TM}} + \frac{1}{m_{TE}}\right) \nabla^2 + \hbar \omega_c -i\gamma_C$, $H_{\pm} = -\frac{\hbar^2}{4} \left(\frac{1}{m_{TM}} - \frac{1}{m_{TE}}\right) \left(\frac{\partial}{\partial x} \mp i \frac{\partial}{\partial y}\right)^2$ and $H_X = -\frac{\hbar^2}{2 m_X}\nabla^2 + \varepsilon_x-i\gamma_X$. Here $E^{\pm}_{i}$ is the cavity mode photon field in the $i$th cavity ($i=1,2$ denoting the two cavities in the double microcavity configuration), and $p^{\pm}_{i}$ is the exciton field in the $i$th quantum well of plus ($+$) and minus ($-$) circular polarization.
 $\hbar \omega_c$ and $\varepsilon_x$ are the $\textbf{k}=0$ energies of the cavity mode and the 1s exciton field, respectively.
%
The two cavities and quantum wells are assumed to be identical and hence the $\textbf{k}=0$ energies of the two cavities are equal.
The cavity photon dispersion is modeled as a parabolic function of momentum, with different masses, $m_{TE}$ and $m_{TM}$, for the TE and TM modes, respectively. $H_{\pm}$ represents the splitting of the TE and TM photon modes that couples the two circular polarization channels. $m_X$ is the 1s heavy-hole exciton mass, which we treat as infinite in the range of interest.  $\gamma_C$ is the cavity loss rate and $\gamma_X$ the exciton dephasing. The coupling strength between excitons and photons in each cavity is $\Omega_X$, and the  photon fields in the two cavities are coupled with strength $\Omega_C$. $R_{\text{pump},1}^{\pm}$ and $R_{\text{pump},2}^{\pm}$ are the effective pump sources in cavity 1 and 2, respectively.

The nonlinear couplings between excitons are discussed in detail in Ref.\ \cite{takayama-etal.02}. They
contain two parts: the Pauli blocking (or phase space filling, PSF) among the excitons' Fermionic constituents (electrons and holes), denoted by $A_{\text{PSF}}$, and the effective interactions $T^{++}$ and $T^{+-}$, in the co-circularly polarized and counter-circularly polarized exciton channels \cite{takayama-etal.02}, respectively. Retardation (quantum memory) effects are neglected due to the quasi-monochromatic excitation conditions.

Diagonalizing the linear part of Eqs.\ (\ref{equ:E-dot-pm-1}) - (\ref{p2.equ}) one finds
two lower polariton branches, LP1 and LP2, and two upper polariton branches, UP1 and UP2. We denote the polariton fields by $\psi_{LP1}^{\pm}$, $\psi_{LP2}^{\pm}$, $\psi_{UP1}^{\pm}$ and $\psi_{UP2}^{\pm}$ respectively. When, as in the experiment, the source frequency is far below the upper polariton branches, the upper polaritons are not excited so that $\psi_{UP1}^{\pm} = \psi_{UP2}^{\pm} =0$.
%
%
The dynamic equations of the lower polaritons in real space are\cite{luk-spie.17}
\begin{eqnarray}
i\hbar \dot{\psi}_{LP1}^{\pm} &=& \tilde{H}_{\psi,1} \psi_{LP1}^{\pm} + \tilde{H}_{\pm,1} \psi^{\mp}_{LP1} +N_{\pm,1} + \tilde{R}_{\pm,1} \label{psirLP1.equ} \\
i\hbar \dot{\psi}_{LP2}^{+} &=& \tilde{H}_{\psi,2} \psi_{LP2}^{+}  +N_{+,2} + \tilde{R}_{+,2} \label{psirLP2.equ}
\end{eqnarray}
with $\tilde{H}_{\psi,1} = - \frac{ \hbar^2}{4} \left( \frac {1} {\tilde{m}_{TM,1}} + \frac {1} {\tilde{m}_{TE,1}} \right)\nabla^2 +\hbar\omega_{0,1}- i \gamma$ and $\tilde{H}_{\psi,2} = - \frac{ \hbar^2}{4} \left( \frac {1} {\tilde{m}_{TM,2}} + \frac {1} {\tilde{m}_{TE,2}} \right)\nabla^2 +\hbar\omega_{0,2}- i \gamma$. Here, $\hbar\omega_{0,1}$ and $\hbar \omega_{0,2}$ are the $\mathbf{k} = 0$ polariton energies for LP1 and LP2, respectively,
 $\tilde{m}_{TE,1}$, $\tilde{m}_{TM,1}$, $\tilde{m}_{TE,2}$, $\tilde{m}_{TM,2}$ are the TE and TM effective LP1 and LP2 polariton masses, respectively,
 and
 $\gamma$ the effective polariton loss rate where we set $\gamma_X = \gamma_C \equiv \gamma$.
The spin-orbit coupling is $\tilde{H}_{\pm,1} = - \frac {\hbar^2} {4} \left( \frac {1} {\tilde{m}_{TM,1}} - \frac {1} {\tilde{m}_{TE,1}} \right)\left( \frac {\partial} {\partial x} \mp i \frac {\partial} {\partial y} \right)^2$.  The quasi-monochromatic source terms at frequency $\omega_p$ are written as $\tilde{R}_{\pm,1}(\mathbf{r},t) = \tilde{R}_{\pm,1}(\mathbf{r})e^{-i\omega_p t}$ and $\tilde{R}_{+,2}(\mathbf{r},t) = \tilde{R}_{+,2}(\mathbf{r})e^{-i\omega_p t}$. We neglect the $\psi_{LP2}^{-}$ terms as we assume that the source on LP2 is ``+'' polarized only, such that $\tilde{R}_{-,2} (\mathbf{r}) = 0$, which leads to $\psi_{LP2}^{-} = 0$.

The nonlinear terms
\begin{eqnarray}
N_{\pm,1} &=& 2 (\tilde{T}^{++}_{12}+\tilde{A}_{PSF,12}) |\psi_{LP2}^{\pm}(\mathbf{r},t)|^2 \psi_{LP1}^{\pm}(\mathbf{r},t) \nonumber \\
&&+ \tilde{T}^{+-}_{12} \Big( |\psi_{LP2}^{\mp}(\mathbf{r},t)|^2 \psi_{LP1}^{\pm}(\mathbf{r},t) + {\psi_{LP2}^{\mp}}^*(\mathbf{r},t) {\psi_{LP1}^{\mp}}(\mathbf{r},t) {\psi_{LP2}^{\pm}}(\mathbf{r},t) \Big) \nonumber \\
&&+ (\tilde{T}^{++}_{11}+\tilde{A}_{PSF,11}) |\psi_{LP1}^{\pm}(\mathbf{r},t)|^2 \psi_{LP1}^{\pm}(\mathbf{r},t) + \tilde{T}^{+-}_{11} |\psi_{LP1}^{\mp}(\mathbf{r},t)|^2 \psi_{LP1}^{\pm}(\mathbf{r},t) \\
N_{+,2} &=& (\tilde{T}^{++}_{22}+\tilde{A}_{PSF,22}) |\psi_{LP2}^{+}(\mathbf{r},t)|^2 \psi_{LP2}^{+}(\mathbf{r},t) + \tilde{T}^{+-}_{12} |\psi_{LP1}^{-}(\mathbf{r},t)|^2 \psi_{LP2}^{+}(\mathbf{r},t) \nonumber \\
&&+ 2(\tilde{T}^{++}_{12}+\tilde{A}_{PSF,12}) \Big( |\psi_{LP1}^{+} (\mathbf{r},t) |^2 \psi_{LP2}^{+} (\mathbf{r},t)\Big) \nonumber \\
&&+ (\tilde{T}^{++}_{12}+\tilde{A}_{PSF,21}) \Big( {\psi_{LP2}^+}^*(\mathbf{r},t) {\psi_{LP1}^+}^2(\mathbf{r},t)\Big)
\end{eqnarray}
arise from the interaction between polaritons, where $\tilde{T}^{++}_{12}$ and $\tilde{T}^{+-}_{12}$ are the effective couplings between two co-circularly and counter-circularly polarized polaritons, one from LP1 and another  from LP2, respectively. Similarly, $\tilde{T}^{++}_{11}$ and $\tilde{T}^{+-}_{11}$ are the effective couplings between two co-circularly and counter-circularly polarized LP1 polaritons, and $\tilde{T}^{++}_{22}$ is the effective coupling between two co-circularly polarized LP2 polaritons. We obtain the polariton interactions from the exciton interactions approximately as $\tilde{T}^{++}_{11} = 2\beta_{11}^2 T^{++}$, $\tilde{T}^{+-}_{11} = 2\beta_{11}^2 T^{+-}$, $\tilde{T}^{++}_{22} = 2\beta_{22}^2 T^{++}$, $\tilde{T}^{++}_{12} = 2\beta_{11}\beta_{22} T^{++}$, $\tilde{T}^{+-}_{12} = 2\beta_{11}\beta_{22} T^{+-}$, $\tilde{A}_{PSF,11} = 4\beta_{PSF,11}^2 A_{PSF}\Omega_X$, $\tilde{A}_{PSF,22} = 4\beta_{PSF,22}^2 A_{PSF}\Omega_X$, $\tilde{A}_{PSF,21} = 4\beta_{PSF,21}^2 A_{PSF} \Omega_X$ and $\tilde{A}_{PSF,12} = 2\beta_{PSF,12}^2 A_{PSF} \Omega_X$. The combinations of Hopfield coefficients are obtained from the projections of the exciton and cavity photon wave functions onto the polariton wavefunctions, and in our case are given by
\begin{eqnarray}
\beta_{11}^2 &=& \frac{\Omega_X^4}{4\left(\Omega_X^2+(\varepsilon_x-\hbar \omega_{0,1}-\frac{\hbar^2 k_{res1}^2}{2M_{LP1}})^2\right)^2} \\
\beta_{22}^2 &=& \frac{\Omega_X^4}{4\left(\Omega_X^2+(\varepsilon_x-\hbar \omega_{0,2})^2\right)^2}\\
\beta_{PSF,11}^2 &=& \frac{\Omega_X^3 (\varepsilon_x-\hbar \omega_{0,1}-\frac{\hbar^2 k_{res1}^2}{2M_{LP1}})}{4\left(\Omega_X^2+(\varepsilon_x-\hbar \omega_{0,1}-\frac{\hbar^2 k_{res1}^2}{2M_{LP1}})^2\right)^2} \\
\beta_{PSF,22}^2 &=& \frac{\Omega_X^3 (\varepsilon_x-\hbar \omega_{0,2})}{4\Big(\Omega_X^2+(\varepsilon_x-\hbar \omega_{0,2})^2\Big)^2}\\
\beta_{PSF,12}^2 &=& \frac{\Omega_X^3 (2\varepsilon_x-\hbar \omega_{0,1}-\frac{\hbar^2 k_{res1}^2}{2M_{LP1}}-\hbar \omega_{0,2})}{4\left(\Omega_X^2+(\varepsilon_x-\hbar \omega_{0,1}-\frac{\hbar^2 k_{res1}^2}{2M_{LP1}})^2\right)\Big(\Omega_X^2+(\varepsilon_x-\hbar \omega_{0,2})^2\Big)} \\
\beta_{PSF,21}^2 &=& \frac{\Omega_X^3 (\varepsilon_x-\hbar \omega_{0,1}-\frac{\hbar^2 k_{res1}^2}{2M_{LP1}})}{4\left(\Omega_X^2+(\varepsilon_x-\hbar \omega_{0,1}-\frac{\hbar^2 k_{res1}^2}{2M_{LP1}})^2\right)\Big(\Omega_X^2+(\varepsilon_x-\hbar \omega_{0,2})^2\Big)}
\end{eqnarray}
where $\frac{1}{M_{LP1}} = \frac {1} {\tilde{m}_{TM,1}} + \frac {1} {\tilde{m}_{TE,1}}$, and $k_{res1}$ is the resonant transverse momentum on the LP1 branch at energy $\hbar \omega_p$.

In the simulations, we use $\tilde{m}_{TM,1} = 7.20\times10^{-5}~m_e$  
and $\tilde{m}_{TE,1} = 1.023 \times m_{TM}$ for LP1, and $\tilde{m}_{TM,2} = 7.57\times10^{-5}~m_e$ 
and $\tilde{m}_{TE,2} = 1.031 \times m_{TM}$ for LP2. Here $m_e$ is the electron mass. The values are obtained from a parabolic fit to the polariton dispersions found from a transfer matrix calculation and experimental data \cite{lafont-etal.17}.
The source energy is chosen at $k=0$ energy of LP2 such that $\Delta_p = \hbar\omega_{0,2}- \hbar \omega_{0,1} = 7.54~\text{meV} $ is taken in agreement with the experimental excitation conditions.
The co-circular exciton-exciton interaction  and phase-space filling parameter are taken to be\cite{ardizzone-etal.13}
$T^{++} = 5.69\times 10^{-3}~\text{meV} \mu \text{m}^{2}$ and $A_{\text{PSF}} = 2.594\times10^{-4}~\mu \text{m}^2$, respectively. The counter-circular exciton-exciton interaction is taken to be $T^{+-} = -T^{++}/3$. We take $\Omega_X = 6.35~\text{meV}$ and $\gamma = 0.2~\text{meV}$.

The solution of the Gross-Pitaevskii equations for the lower polariton branches, Eq.~(\ref{psirLP1.equ}) and (\ref{psirLP2.equ}), allows us to analyze the pseudo-spin vector elements (Stokes parameters) in real space.
The normalized (superscript $n$) pseudo-spin vector elements in real space  are
$ S^{(n)}_1 ({\bf r} ) = ( | \psi^{x}_{LP1} ({\bf r} )  |^2   -   | \psi^{y}_{LP1} ({\bf r} )  |^2 )/ S_0 ({\bf r} )$
(where the subscripts $x$ and $y$ refer to linear polarization)
and
$ S^{(n)}_3 ({\bf r} ) = ( | \psi^{+}_{LP1} ({\bf r} )  |^2   -   | \psi^{-}_{LP1} ({\bf r} )  |^2 )/ S_0 ({\bf r} )$
with
$ S_0 ({\bf r} ) =  | \psi^{+}_{LP1} ({\bf r} )  |^2   +  | \psi^{-}_{LP1} ({\bf r} )  |^2 $.

\begin{figure}
\includegraphics[scale=0.8]{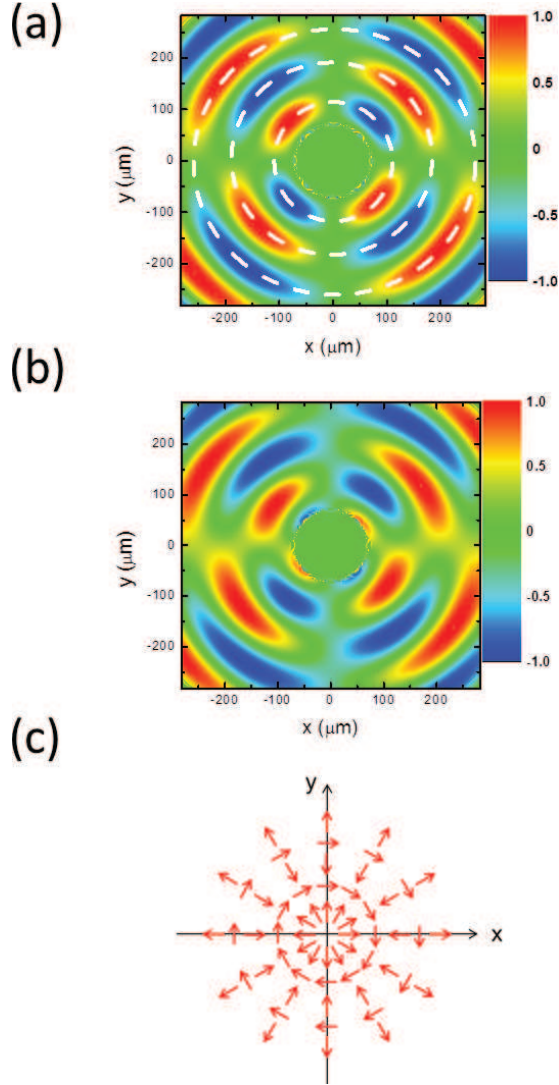}
\caption{
(a) Numerical results for the polarization  $ S_1 ({\bf r})$, with red (blue) indicating  predominantly X (Y)  polarization,  at low excitation powers, in agreement with Ref.\ \cite{kammann-etal.12}. In (b), a higher excitation power is used, resulting in a rotation/twisting of the polarization pattern. (c) A schematic representation of the spin current showing radially alternating directions, separated by areas of (very small) tangential spin currents (these areas are shown as dashed white circles in (a)).
For clarity, all vectors are normalized to the same length.
}
\label{fig:r-space-texture}
\end{figure}

The numerical simulation results at steady state of Eqs.~(\ref{psirLP1.equ}) and (\ref{psirLP2.equ}) are shown in Fig.~\ref{fig:r-space-texture} at different excitation power. The source is circularly (``+'') polarized with Gaussian beam profile. The steady-state linear polarization texture in real space, $ S_1 ({\bf r})$,  at low excitation power exhibits the feature shown in Ref.\ \cite{kammann-etal.12}, see Fig.\ \ref{fig:r-space-texture}a. It is also instructive to compute the spin current, defined as
 \begin{equation}
\mathbf{J} = \frac{\hbar^2}{m_{LP1}}\left[ \hat{r} \text{Im}\left( \psi_{LP1}^{+ *} \frac{\partial}{\partial r} \psi_{LP1}^{+} -\psi_{LP1}^{- *} \frac{\partial}{\partial r} \psi_{LP1}^{-} \right) - \hat{\phi} \frac{2}{r} |\psi_{LP1}^{-}|^2\right] \label{spincurrent.equ}
\end{equation}
where  $\frac{1}{m_{LP1}} = \frac {1} {\tilde{m}_{TM,1}} - \frac {1} {\tilde{m}_{TE,1}}$. For clarity, we display the result in a schematic fashion in Fig.\ \ref{fig:r-space-texture}c and omit the spin current texture due to phase oscillations of $\psi^{+}({\bf r})$ within the pump profile in this figure.
 Pumping spin-up  polaritons close to the origin  leads to a radial outward flow of predominantly spin-up polaritons due to the polariton's ballistic motion. However, the spin-orbit coupling results in a reversal of the radial flow in an adjacent ring of predominantly spin-down polaritons, the two regions being separated by a narrow ring of tangential spin current (whose magnitude is very small). Increasing the pump source leads to a deformation of the polarization texture as shown in Fig.\ \ref{fig:r-space-texture}b.



\section{Far-field spin/polarization texture rotation}

In addition to the real space formulation, we also analyze the results in momentum space.
We perform a Fourier transform of the polariton wave functions and obtain
$\psi^{\pm}_{LP1}(\textbf{k})$, which determines the off-axis emission related to the occupation of the elastic circle,
and define the Stokes parameters, i.e.\ the pseudo-spin, in terms of the LP1 wave functions.
 The $S_1$ component of the k-dependent pseudo-spin vector is, without conventional normalization denominator,
 $ S_1 ({\bf k} ) =  | \psi^x_{LP1} ({\bf k} )  |^2   -   | \psi^y_{LP1} ({\bf k} )  |^2 $.
  With that normalization included, it reads
 $ S^{(n)}_1 ({\bf k} ) = S_1 ({\bf k} ) / [ | \psi^x_{LP1} ({\bf k} )  |^2   +   | \psi^y_{LP1} ({\bf k} )  |^2 ]$. Alternatively, we can write
 $ S_1 ({\bf k} ) = 2 {\rm Re} [  \psi_{LP1}^{- \ast} ({\bf k})    \psi^{+}_{LP1} ({\bf k}) ]$,
 and furthermore,
 $ S_2 ({\bf k} ) = - 2 {\rm Im} [  \psi_{LP1}^{- \ast} ({\bf k})    \psi^{+}_{LP1} ({\bf k}) ]$
 and
 $ S_3 ({\bf k} ) =  | \psi^+_{LP1} ({\bf k} )  |^2   -   | \psi^-_{LP1} ({\bf k} )  |^2 $.

We note that, because of its definition in terms of squared wave functions,
 the pseudo-spin vector in wave vector space, that enters the torque equation discussed above, is not the spatial Fourier transform of the configuration space pseudo-spin vector, so e.g.\  the  1-component
 $ S_1 ({\bf r})$  without normalization denominator is not the Fourier transform of
 $ S_1 ({\bf k} )  $.  The normalized $S^{(n)}_1$     ($S^{(n)}_3$ ) component provides the degree of linear (circular) polarization.

\begin{figure}
\includegraphics[scale=0.7]{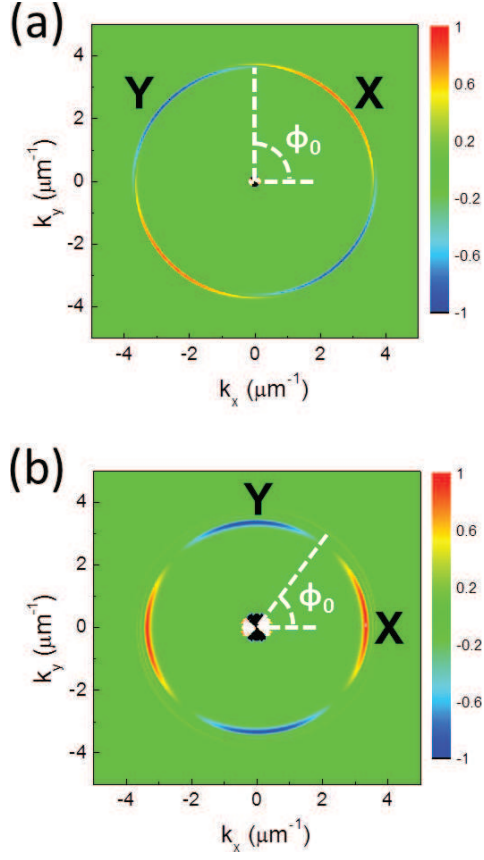}
\caption{
 Numerical results for the polarization  $ S_1 ({\bf k})$, with red (blue) indicating  predominantly X (Y)  polarization for a Gaussian pump pulse with diameter 50 $\mu$m (FWHM in intensity).
Here, the data are normalized such that the maximum (minimum) value of $ S_1 ({\bf k})$  on the elastic circles is +1 (-1).
The function  $S_1 ({\bf k})$ contains information about the linear polarization seen in the far field emission.
  Low excitation power shown in (a), high power with an exciton density of $164~\mu$m$^{-2}$ shown in (b).
 The nonlinear OSHE clockwise rotation of about 37$^{\circ}$ is clearly seen.
}
\label{fig:k-space-texture}
\end{figure}

Using a low-intensity Gaussian beam profile with FWHM (in intensity) of 50$\mu$m
for the circularly (``+'') polarized source, the steady-state linear polarization texture in momentum space, $S_1(\mathbf{k})$, is shown in
Fig.\ \ref{fig:k-space-texture}a. It exhibits the linear OSHE effect, in agreement with
Fig.\ \ref{fig:steady-state-torque}.
 Since quasi-monochromatic pumping occupies only states on the (hardly distinguishable) elastic circles of the TE and TM modes, the polarization texture is restricted to the vicinity of the circles.
Increasing the pump intensity yields a rotation of the polarization textures as shown in Fig.\ \ref{fig:k-space-texture}b, again in agreement with the expectations of Fig.\
 \ref{fig:steady-state-torque}d.

In order to quantitatively determine the rotation of the far-field spin/polarization texture,
   we average each term entering the k-dependent Stokes parameter
along the radial direction over the ring that includes the elastic circles of both the TE and TM modes,
and denote the average as $\langle  | \psi^{x/y}_{LP1} ({\bf k} )  |^2  \rangle$,
 which is a function of the polar angle $\phi_k$, which from now on we denote simply by $\phi$. We can now use these radial averages in the definition of the Stokes parameters, in particular $S_1$, and obtain a Stokes parameter, denoted by $ \bar{S}_1 (\phi)$,
 that depends only on $\phi$.
We could see from the numerical results presented in Fig.~\ref{fig:k-space-texture} that  $ \bar{S}_1 (\phi)$ (and  the normalized counterpart $ \bar{S}^n_1 (\phi)$)
 displays a $\cos(2 \phi + \Theta)$ dependence, where $\Theta$ is an offset determined by the pump field characteristics such
 as pump power. Therefore a rotation of the far-field pattern  with increasing power is equivalent to a power-dependent change of  $\Theta$.
To quantify such a rotation, we denote by $\phi_0$ the angle at which $ \bar{S}_1 (\phi)$ first vanishes in the range between $0$ to $2\pi$, i.e.\
$ \bar{S}_1 (\phi_0) \sim \cos(2 \phi_0 + \Theta) =0$, as shown in Fig.~\ref{fig:k-space-texture}.


It is desirable to have
 a simple analytical approximation for the changes of $\Theta$. To this end, we proceed as follows. We consider the nonlinear coupling effect of polaritons on  $S_1(k)$, which is assumed to have LP1 polaritons only on the elastic circles (with $k\ne0$). The nonlinear effect is expected to be dominated by the polaritons on LP2 with the same polarization as the source, $\psi^{+}_{LP2}$. Therefore, as an approximation that can be validated through comparison with our full numerical solutions, we ignore the contributions from $\psi^{\pm}_{LP1}$ to the density in the polariton interaction term $N_{\pm,1}$ in Eq.~(\ref{psirLP1.equ}). Under this approximation, the $\mathbf{k}$-space version of Eq.~(\ref{psirLP1.equ}) is
\begin{eqnarray}
i\hbar \dot{\psi}^{\pm}_{LP1}(\mathbf{k},t) &=& \Big(\frac{\hbar^2 k^2}{2 \tilde{M}_{1}} + \hbar \omega_{0,1} - i\gamma\Big) \psi^{\pm}_{LP1}(\mathbf{k},t) + \frac{\hbar^2 k^2}{2\tilde{m}_{1}} e^{\mp 2i\phi} \psi^{\mp}_{LP1} (\mathbf{k},t) \nonumber \\
&& + \int \frac{d\mathbf{k'}}{(2\pi)^2} \frac{d\mathbf{k''}}{(2\pi)^2} W_{\pm} {\psi^{+}_{LP2}}^{*}(\mathbf{k'}+\mathbf{k''}-\mathbf{k}) \psi^{+}_{LP2}(\mathbf{k'}) \psi^{\pm}_{LP1} (\mathbf{k''}) \nonumber \\
&&+ \tilde{R}_{\pm,1}(k,t), \label{psikLP1.equ}
\end{eqnarray}
with $W^+ = 2 (\tilde{T}^{++}_{12}+\tilde{A}_{PSF,12})$ and $W^- = \tilde{T}^{+-}_{12}$. Furthermore, $\frac{1}{\tilde{M}_{1}} = \frac{1}{2} \left( \frac{1}{\tilde{m}_{TM,1}} + \frac{1}{\tilde{m}_{TE,1}} \right)$ and $\frac{1}{\tilde{m}_{1}} = \frac{1}{2} \left( \frac{1}{\tilde{m}_{TM,1}} - \frac{1}{\tilde{m}_{TE,1}} \right)$, and  $\tilde{R}_{-,1} = 0$ when the source is ``+'' polarized. Since the LP2 density is narrowly distributed around $\mathbf{k} = 0$, we approximate the $\psi^{+}_{LP2}(\mathbf{k})$ in the density term by $\psi^{+}_{LP2}(\mathbf{k}) \approx \delta(\mathbf{k})\int d\mathbf{k'} \psi^{+}_{LP2}(\mathbf{k'}) = \delta(\mathbf{k}) (2\pi)^2 \psi^{+}_{LP2}(r=0)$. Substituting this into Eq.~(\ref{psikLP1.equ}), we have
\begin{eqnarray}
i\hbar \dot{\psi}^{\pm}_{LP1}(\mathbf{k},t) &=& \Big(\frac{\hbar^2 k^2}{2 \tilde{M}_{1}} + \hbar \omega_{0,2} + W^{\pm}|\psi^{+}_{LP2}(r=0)|^2 - i\gamma\Big) \psi^{\pm}_{LP1}(\mathbf{k},t) \nonumber \\
&& + \frac{\hbar^2 k^2}{2\tilde{m}_{1}} e^{\mp 2i\phi} \psi^{\mp}_{LP1} (\mathbf{k},t) + \tilde{R}_{+,1}(k,t) \label{psik_simp.equ}
\end{eqnarray}
The stationary solutions of Eq.~(\ref{psik_simp.equ}) are
\begin{eqnarray}
\psi^{+}_{LP1}(k) &=& \Big( \Delta_p - \frac{\hbar^2 k^2}{2\tilde{M}_{1}} - \Delta E_{-} + i\gamma \Big) \frac{\tilde{R}_{+,1}(k)}{D} e^{-i \Delta_p t} \label{equ:psipkss}\\
\psi^{-}_{LP1}(k) &=& \frac{\hbar^2 k^2}{2 \tilde{m}_{1}} \frac{\tilde{R}_{+,1}(k)}{D} e^{2i\phi} e^{-i \Delta_p t} \label{equ:psimkss}
\end{eqnarray}
where
\begin{eqnarray}
D &=& \Big(\Delta_p - \frac{\hbar^2 k^2}{2\tilde{M}_{1}} - \Delta E_+ + i\gamma \Big) \Big(\Delta_p - \frac{\hbar^2 k^2}{2\tilde{M}_{1}} - \Delta E_{-} + i\gamma \Big) \nonumber \\ &&- \Big(\frac{\hbar^2 k^2}{2 \tilde{m}_{1}} \Big)^2
\end{eqnarray}
and the energy shifts are $\Delta E_+ = 2(\tilde{T}^{++}_{12}+\tilde{A}_{PSF,12})|\psi^{+}_{LP2}(r=0)|^2$ and $\Delta E_- = \tilde{T}^{+-}_{12}|\psi^{+}_{LP2}(r=0)|^2$.

As $ S_1 ({\bf k} ) = 2 {\rm Re} [  \psi_{LP1}^{- \ast} ({\bf k})    \psi^{+}_{LP1} ({\bf k}) ]$ is proportional to $\cos(2\phi + \Theta_k)$, we have \begin{eqnarray}
\Theta_k &=& arg(\psi^{+}_{LP1} ({\bf k})) - arg(\psi_{LP1}^{-} ({\bf k})) \nonumber \\
&=& -\arctan \Big(\frac{\gamma}{\Delta_p - \frac{\hbar^2 k^2}{2\tilde{M}_{1}}-\Delta E_{-}} \Big) \label{equ:thetak}
\end{eqnarray}
from the arguments of the stationary solutions in Eqs.~(\ref{equ:psipkss}) and (\ref{equ:psimkss}).

 When we average $S_1(\mathbf{k})$ along the radial direction,
 the observed angular shift becomes
\begin{equation}
\Delta \Theta = \Theta_{k_{\text{nonlin}}} - \Theta_{k_{\text{lin}}} \label{equ:deltatheta}
\end{equation}
where $k_{\text{lin}}$ and $k_{\text{nonlin}}$ are the resonant $k$-values in the linear and non-linear case, respectively.
We find that, to a very good approximation, for Gaussian beam profiles
the linear phase is  $\pi/2$. In the non-linear case, $\psi^{+}_{LP1}(k)$ is approximately resonant at
\begin{equation}
\Delta_p - \frac{\hbar^2 k^2_{\text{nonlin}}}{2\tilde{M}_{1}} \approx \Delta E_+ \label{resonance-cond.equ}
\end{equation}
giving the angular shift as
\begin{equation}
\Delta \Theta = \arctan \Big(\frac{(2(\tilde{T}^{++}_{12}+\tilde{A}_{PSF,12}) - \tilde{T}^{+-}_{12})|\psi^{+}_{LP2}(r=0)|^2}{\gamma} \Big),
\end{equation}
which is a consequence of excitonic interaction.
Then, $\phi_0$ is given by
\begin{eqnarray}
\phi_0 &=& \frac{\pi}{4} - \frac{\Theta_{k_{\text{nonlin}}}}{2} \nonumber\\
& \approx & \frac{1}{2} \left(\pi -
 \arctan \left[ ((2 (\tilde{T}^{++}_{12}+\tilde{A}_{PSF,12}) - \tilde{T}^{+-}_{12} ) | \psi^{+}_{LP2} (\textbf{r}=0) |^2 ) / \gamma \right]
 \right)
 \label{equ:angular-shift-simple-model}
\end{eqnarray}
where the resonance condition  Eq.~(\ref{resonance-cond.equ}) has been used.
 It provides an analytical estimate of the angular rotation of the polarization texture as a function of the LP2 density $| \psi^{+}_{LP2} (\textbf{r}=0) |^2 $ which is created by the ``+'' polarized pump source. Note that this analytical model  is valid for infinite spot size of the beam. We also note that the angular rotation in Eq.\ (\ref{equ:angular-shift-simple-model}) changes sign if we use a ``-'' instead of  ``+'' polarized source.

The expression in Eq.\ (\ref{equ:angular-shift-simple-model}) allows us to identify the pump-induced $B_3$ component of the effective magnetic field, which was left as a parameter in the  above discussion of the pseudo-spin model, as
\begin{equation} \label{equ:B3-given-microscopically}
B_3 = (2 ( \tilde{T}^{++} + \tilde{A}_{PSF} )- \tilde{T}^{+-} ) | \psi^{+}_{LP2} (\textbf{r}=0) |^2
\end{equation}
From a microscopic point of view, the control of $B_3$ is based here on the interaction between LP1 and LP2 polaritons, with the LP2 population giving rise to Coulombic and phase-space filling
 energy shifts of the two LP1 polariton spin states. Since $\tilde{T}^{++}$ and $\tilde{A}_{PSF}$
 are positive and $\tilde{T}^{+-}$ negative, the difference between the shifts due to co-circularly and counter-circularly polarized states leads to a reinforcement, rather than a cancellation, of the nonlinear OSHE rotation. We note again that the analytic expression for $B_3$ in Eq.\ \ref{equ:B3-given-microscopically} is valid for infinite beam spot size.

One advantage of the microscopic theory developed here is that is allows for a quantitative estimate of the rotation of the far-field spin/polarization texture as a function of the incident power.
In \cite{lafont-etal.17}, the simple pseudo-spin model was used to analyze experimental data, but since $B_3$ is a phenomenological  parameter a quantitative power estimate cannot be obtained from that model. In the following,
we show how the present theory can give that estimate for arbitrary beam spot sizes.
%
%
Starting from the condition of energy conservation,
 we have derived the following
 relation
\begin{equation}
\label{equ:p_inc-vs-exciton-density}
P_{inc} = \left( \frac {\pi d_{FWHM}^2} { 4 {\rm ln}2} \right) \left( \frac {\varepsilon_x \gamma_{rec}} {\hbar ( 1 - | r |^2 - \frac {n_{tran}} {n_{inc}} | t |^2 )} \right) \sum\limits_{j=1}^{N} |p_j|^2,
\end{equation}
Here, $|p_j|^2$ is the exciton density in the $j$-th quantum well,
$\gamma_{rec}$ the electron-hole recombination rate (describing decay of the exciton density via radiative and non-radiative recombination),
$|r|^2$ is the reflectance of the doublecavity system, $|t|^2$ its transmittance, $n_{inc}$ and $n_{tran}$ are the refractive indexes of the materials on the side of the incident and transmitted beam, respectively. From numerical transfer matrix simulations we obtain $|r|^2 = 0.391$, $|t|^2 = 0.014$ with $n_{inc} = 1$ and $n_{tran} = 3.59$.
$d_{FWHM}$ is the full-width at half-maximum of the pump intensity with a Gaussian profile, which in the experiment is estimated to be on the order of $50~\mu m$. Since the OSHE rotation is a consequence of excitonic interaction, it is beneficial to study the
rotation as a function of exciton density, not the polariton density (because the photonic component of the polaritons does not contribute to the rotation).
 From the theory outlined above, the pump-induced exciton density of each quantum well is given by
\begin{equation}
|p_j|^2  \equiv |p|^2 =
\beta_{22} |\psi^{+}_{LP2}(r=0)|^2  \label{poltoexciton.equ}
\end{equation}
which we approximate to be the same for all  $j$ (the small differences of the photonic environment for the different quantum wells can be ignored for the present approximate power estimate). The factor $\beta_{22}$ is given in Eq.\ (10).
%
An estimate of the recombination time in our sample can be obtained from Refs.\ \cite{bajoni-etal.06,bastard.89}. For a 20~nm wide (almost bulk-like) quantum well, Ref.\ \cite{bajoni-etal.06} reports a 700~ps radiative exciton lifetime. The difference between lifetimes in bulk and two-dimensional systems is about a factor of 4, see Ref.\ \cite{bastard.89}. Since the quantum wells in our sample are  thin (7nm) and have large barriers, and are therefore closer to the two-dimensional limit, we estimate our radiative lifetime to be approximately 700~ps/4,
or between 100 and 200~ps.

Applying these considerations to the case of Ref.\ \cite{lafont-etal.17}, we find that
for a spot size of 50~$\mu$m
and a lifetime (including radiative and non-radiative) of
 $\tau_{rec} = 92~$ps,
 a power of 40~mW corresponds to an exciton density of 120~$\mu$m$^{-2}$.



\begin{figure}
\includegraphics[scale=0.8]{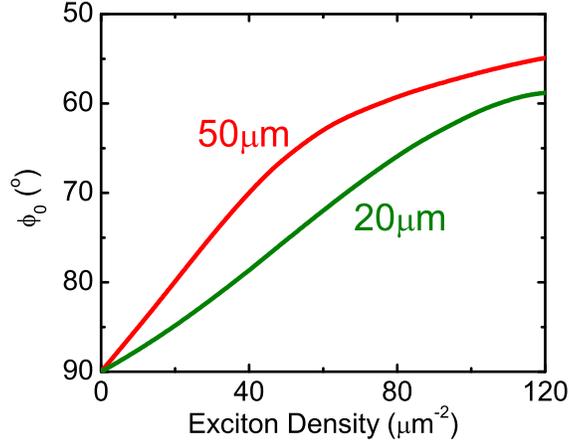}
\caption{
 Theoretical angular orientation (zero-crossings) $\phi_0$ of the radial average, $ \bar{S}^{(n)}_1$,
 from numerical solutions of the 2-dimensional spinor-polariton Gross-Pitaevskii equation for finite spot diameters $d_{FWHM}$, indicated in the plot.
 The horizontal axis represents the exciton density, which is a fraction of the LP2 polariton density,
 $|p|^2 = \beta_{22} |\psi^{+}_{LP2}(r=0)|^2 $ (see text).
}
\label{fig:phi_0}
\end{figure}

 In Fig.\ \ref{fig:phi_0} we show the orientation $ \phi_0 $
 %
obtained from numerical solutions of the polaritonic Gross-Pitaevskii
equation with finite spot size.
In Fig.\ \ref{fig:phi_0},
for fixed exciton density, increasing the spot size yields a larger rotation of the OSHE pattern.  We restrict ourselves here to Gaussian pulse profiles and moderate  densities. A detailed analysis of
the OSHE rotation caused by other pump pulse profiles and over a larger range of  densities will be given elsewhere. We also note that
in the limit of infinite spot size of a Gaussian beam, the numerical results approach the simple analytical model, Eq.\ (\ref{equ:angular-shift-simple-model}).



\section{Conclusion}

In conclusion, we have shown that Coulombic interactions between polaritons on the two lower branches, LP1 and LP2, of a semiconductor double microcavity can be used to control the spin/polarization patterns in real space (near field) and momentum space (far field). Using a  double-cavity spinor Gross-Pitaevskii equation, we have found in particular that the far field pattern, which determines the anisotropic ballistic polariton transport,
 can undergo a simple rotation as a function of increasing LP2 polariton density.
This is in agreement with recent experimental results.
The rotation depends on the spot size of the excitation beam, and, in the limit of infinite spot size, approaches a simple analytical model, Eq.\ (\ref{equ:angular-shift-simple-model}).
The microscopic theory presented here also allows for an identification of the $B_3$ component of a phenomenological pseudo-spin model,
which is responsible for the far-field rotation. The $B_3$ component is shown to be related to the excitonic T-matrix in the co- and counter-circularly polarized interaction channel as well as the LP2 polariton density,
Eq.\ (\ref{equ:B3-given-microscopically}). Further studies of the dependence of the spin/polarization texture control on the spatial pump profile and over a wide range of intensities,
reaching to the OPO threshold, will be interesting extensions of the present work.

\section{Acknowledgement}

We gratefully acknowledge financial support from NSF under grant ECCS-1406673, TRIF SEOS, and the German DFG
(TRR142, SCHU 1980/5, Heisenberg program).
We thank Kyle Gag for helpful discussions.





\end{document}